\documentclass{aa}
\usepackage{graphicx}
\usepackage{latexsym}
\usepackage{txfonts}
\usepackage[figuresright]{rotating}
\begin{document}
\renewcommand{\labelitemi}{-}

\title{Size dependence of solar X-ray flare properties}
\author{Marina Battaglia
 \and Paolo C. Grigis 
  \and Arnold O. Benz}
\institute{Institute of Astronomy, ETH Zentrum SEC, CH-8092 Z\"urich, Switzerland}
\date{Received 8 March 2005 / Accepted 24 April 2005}

\abstract{Non-thermal and thermal parameters of 85 solar flares of GOES
class B1 to M6 (background subtracted classes A1 to M6) have been compared
to each other. The hard X-ray flux has been  measured by RHESSI and a
spectral fitting provided flux and spectral index of the non-thermal
emission, as well as temperature and emission measure of the thermal
emission. The soft X-ray flux was taken from GOES measurements. 
We find a linear correlation in a double logarithmic plot between the
non-thermal flux and the spectral index. The higher the acceleration rate of a
flare, the harder the non-thermal electron distribution. The relation is similar to the
one found by a comparison of the same parameters from several sub-peaks of a single
flare. Thus small flares behave like small subpeaks of large flares. 
Thermal flare properties such as temperature, emission measure and the soft
X-ray flux also correlate with peak non-thermal flux. A large 
non-thermal peak flux entails an enhancement in both thermal parameters. 
The relation between spectral index and the non-thermal flux is an intrinsic
feature of the particle acceleration process, depending on flare size. This
property affects the reported frequency distribution of flare energies. 

 }

\maketitle


\section{Introduction}
Do small flares differ from large flares? In the simplest flare scenarios
small flares are just
scaled down versions of large flares. This is predicted explicitly in
theories that envision flares as a kind of avalanche process, consisting of
many small elements (Lu \& Hamilton \cite{Lu91}; Lu et al. \cite{Lu93}). On the other hand,
one could speculate that a flare changes the environment of an active region
and, in its course, alters the properties of the elementary events. Such an effect would be a
feedback effect, thus a
secondary phenomenon. However, big flares could be genuinely different
because the initial conditions that lead to them may be more demanding on
some plasma parameters that finally not only make them bigger but qualitatively dissimilar.

Kahler (\cite{Kahler82a}) noted that there is a statistical correlation between solar
flare energy release and the observation of certain flare manifestations. 
This so-called `Big Flare Syndrome' states that large flares tend
to be associated with phenomena that may not be directly related to each
other, such as solar protons (Kahler \cite{Kahler82b}), Type II and IV radio bursts
(Kahler \cite{Kahler82b}), decimeter radio emissions (Simnett \& Benz
\cite{Simnett86}), coronal mass ejections
(Dougherty et al. \cite{Doug02}), and white light flares (Matthews et
al. \cite{Matthews03}). A simple interpretation of the Big Flare Syndrome proposes that
the various signatures just get above the threshold for observations in large
events (Kahler 1982a). However, some radiations do not increase linearly with
flare size. Such a behavior was e.g. recently reported for the energy in
decimeter type IV bursts, increasing with about the fifth power of the GOES
soft X-ray flux (Benz et al. \cite{Benz05}).

A quantitative difference between small and large flares was found in thermal
soft X-ray emission. Feldman et al. (\cite{Feldman96}) report an increase of temperature
$T$ with the flare soft X-ray emission, defined by the GOES peak flux $F_\mathrm{G}$ at
1-8 \AA\ (1.55-12.4~keV). The average relation is approximately 
\begin{equation} \label{feldmaneq}
F_\mathrm{G}(T)=3.5\cdot 10^{0.185 T - 9.0}, 
\end{equation}
where the flux is in Wm$^{-2}$ and the
temperature in units of 10$^6$ K. They also find a correlation between
  emission measure and temperature which has been explained theoretically by
  Shibata \& Yokoyama (\cite{Shibata}) based on a balance between magnetic reconnection
  and conductive cooling. Feldman et al. conclude that if a flare is the collection of
subresolution events, the plasma properties of the events occurring during
peak emission of large flares and small flares must be different.

Here we ask whether the characteristics of {\it hard} X-rays emitted by the
non-thermal electrons accelerated in solar flares are different in small and
large flares. Except for the rare thermal flares, the energy distribution of
hard X-ray flare photons follows a power-law. Its index $\gamma$ usually
starts with a high value ('soft' radiation), reaches a minimum at peak flux
and increases at the end (as noticed already by Parks \& Winckler \cite{Parks69}). This
'soft-hard-soft' (SHS) behavior is also found in sub-peaks and has recently
been analyzed quantitatively for the first time by Grigis \& Benz
(\cite{Grigis04}, henceforth GB04) using RHESSI observations.
The SHS evolution of individual flares and subpeaks indicates a change in the acceleration
process in the course of a flare. It can be visualized roughly by the
non-thermal spectrum in log-log representation moving down and up at high
energies with a fixed point in the flux-index plane, at an energy of 6.5 -
12.5~keV (average 9~keV; GB04). Although this 'pivot' point
may be virtual (i.e. an extrapolation beyond measured data), its location is a
characteristic of the electron acceleration process and the subsequent diffusion in energy. 

For all rise and
decay phases in several subpeaks of 24 flares, GB04 find the relation
\begin{equation}
\gamma  = A F_{35}^{-\alpha} ,
\label{gammafluxfunc}
\end{equation}
where $F_{35}$ is the fitted non-thermal flux at 35~keV, and $\alpha=0.197\pm
0.003$. 
Does the same relation hold between the flux and
spectral index at peak time for flares of different size? 
Another question is whether the flares of different size behave
in the average like the subpeaks of one flare. X-ray data of a large number
of flares, small and large, have been registered by the Ramaty High Energy
Solar Spectroscopic Imager (RHESSI, Lin et al. \cite{Lin02}). 
Section \ref{eventselection} describes how an equal number of
flares in each decade of peak GOES flux was selected. In
Sect.~\ref{results}, the
 characteristic values of
thermal and non-thermal emissions at peak flux are compared as a function of
flare size. The results are
discussed and compared to previous work in Sect.~\ref{discussion}, and
conclusions drawn in Sect.~\ref{summary}.


\section{Event selection and data reduction} \label{eventselection}
The RHESSI satellite observes the full Sun from 3~keV to 17~MeV since February
2002. In this study, the high spectral resolution (1 keV) of RHESSI's
germanium detectors is utilized to study the hard X-ray flare emission below 300 keV. This 
energy range is dominated by a thermal part at the lower end, usually
superposed by a non-thermal part with a power-law spectrum at higher
energies. The high spectral resolution allows for the first time to separate
and to analyze the non-thermal component where it comprises most photons in a
large number of flares. The same analysis also yields information on the
thermal X-ray emission in the 3--20~keV range. It is supplemented by soft
X-ray observations from the GOES satellite (Garcia
\cite{Garcia94}). 

We selected 100 well observed flares from GOES class B1 to X1 from the time
between RHESSI launch and April 5th
 2004, using the RHESSI Experimental Data Center (HEDC, Saint-Hilaire et
 al. 2002) at ETH Z\"urich. The selection
process aimed at having a statistically representative sample of flares over
the whole time period and from different active regions. \\

\subsection{Data selection}

We took all events with an image in HEDC, giving 
their location on the 
solar disc, and present in the 
RHESSI-flare-list of April 5th, 2004.
This ensured that only solar events were selected and yielded a list of 6039
flares. As we wanted to arrange the flares
according to their GOES class, flares without or with bad GOES data in the
corresponding time interval where removed from the list. This second step ended
with a list of 5871 flares.

To get a uniform distribution of events in flare size, flares where sorted
according to their GOES peak flux (without
background subtraction). The flux range B1 to X1 was divided into 10 equally
 wide bins on a logarithmic scale.
 For each of the 10 bins, 10 events where randomly chosen. 
 A flare had to fulfill the following criteria to be acceptable:
\begin{enumerate}
 \item The start of the flare and its peak hard X-ray emission
  had to be well observed. This should guarantee that we did not miss any
  interesting parts and allow a good background
  selection.  
\item The attenuation state had to be constant for a period of 4 minutes
  centered at hard X-ray peak time.
 \item The front decimation weight had to be less or equal to 2, to better
 study the thermal emission. There was an exception from
 this rule for GOES classes M5 to X1, because in this bin there were just 8
 events. Thus every single flare that fulfilled
 the first 2 criteria has been 
 taken to have a sample as large as possible.  
\item For flares in GOES class B1 to B9, there had to be no attenuation and decimation,
  because of the weak emission of these flares and the low
  count rates. 
\item There must be no enhanced flux of charged particles in the satellite environment a few
  minutes before and after peak time.  
\end{enumerate}
We additionally discarded small flares occurring during the decay phase
of a larger event since in such cases it is not possible to separate the
emissions of the two events. 
 
For each of the selected flares we performed a spectral analysis. 
We chose the longest time interval around flare peak time without gaps and with
RHESSI in sunlight, and we created a series of spectra with a time bin with of
one RHESSI rotation period (approximately 4 seconds). 
The energy-bin width was chosen  fine enough to resolve the thermal and
non-thermal spectra, but not so fine that statistical errors became too large. In our case that meant
a binning of 1 keV from 3--50~keV (with exception of the range 6--12~keV for the
larger flares (C3--X1), where the binning was 0.3~keV), and a larger binning
above 50 keV. With these bin
widths, a clear separation between the thermal and
non-thermal spectrum is possible. For smaller flares, a wider binning was
applied for energies of
20~keV and higher as there is less flare emission above
that energy. 
Only the data from the front segments, without
detector 2 and 7, where used (in case the transmitter was active during the chosen time
interval, detector 8 was also omitted). 

For 7 events, no spectrum files could be produced. One event had the lifetime
below 90$\%$ at peak time and therefore was not acceptable for spectroscopy. 
This left us with 92 events. 

\subsubsection{Selection effect for small flares} \label{selectioneffect}
The RHESSI flare list reports events automatically flagged by the software
when the emission in the 12--25~keV band increases. This energy range contains
some non-thermal emission of microflares (Benz $\&$ Grigis
\cite{Benz02}; Krucker et al. \cite{Krucker02}). However, weak GOES events tend to be lost in the RHESSI
background in this energy range. Therefore, the RHESSI flare
list misses many soft X-ray events of lower GOES class. We have tried to
compensate for it by choosing manually 14 flares from the observing summary
light curves that did not have a flare flag. For all 
of those events, the non-thermal emission was however too small to perform a
spectral fitting in such a way that the event could be used meaningfully in the further
analysis. Thus there remains a selection effect: for low GOES classes, events
with large 12--25~keV flux are preferentially selected. Comparisons between
non-thermal emission and thermal emission (GOES class), where this effect may
play a role, must be treated with special attention. We will point them out in
the discussion (Sect.~\ref{discussion}). The influence of the  selection
effect on the relation between the non-thermal parameters is shown in Fig.~\ref{fluxvsgamma}.

\subsection{Background subtraction and peak time selection}
For each event we subtracted the background and performed a spectral fitting at
the time of maximum emission in the hardest observable peak. As the spectral
index changes with time, we chose a fitting time-interval of only 4 s to
have a value for the instantaneous spectral index, rather than one averaged over time.
The fitting has been performed using the SPEX package (Schwartz
\cite{Schwartz96}; Smith et al. \cite{Smith02}). 
In total, 3 time intervals have been fitted: One at peak time, and one
immediately before and after the peak time interval to compare the fittings
and see whether they were plausible. For the later analysis only the spectrum
at peak time has been used. 
\subsection{Spectral fitting}
SPEX transforms a model photon spectrum
into a model count spectrum via the spectral response matrix (SRM) and compares
it to the observed count spectrum, iteratively adjusting the model parameters
until a local minimum in $\chi^2$ is found. 

\begin{figure} 
\resizebox{\hsize}{!}{\includegraphics{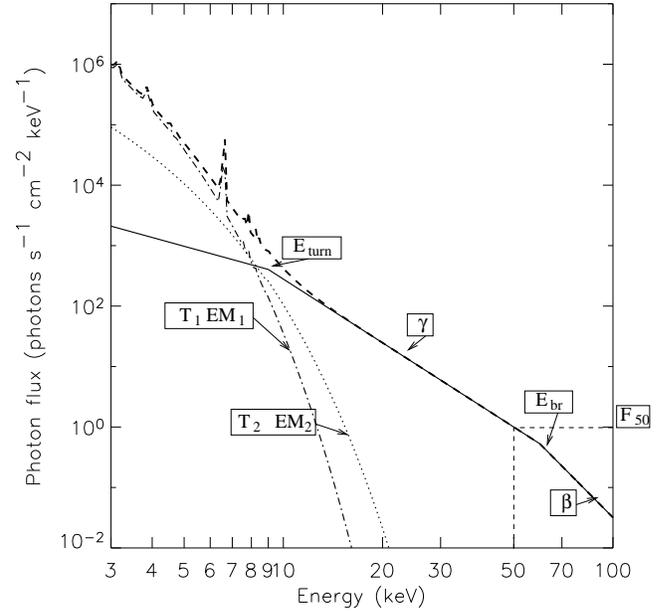}} 
\caption{\small Illustration of the fitting model and the 9 free
  parameters discussed in the text. A
  non-thermal component (solid line) and two thermal components (dotted and
  dash-dotted) are shown.} \label{f_multi}
\end{figure}

We used a spectral model that consisted of 2 isothermal components, 
each given by its temperature $T_1$ and
$T_2$ as well as its 
emission measure $EM_1$ and $EM_2$ respectively, and a non-thermal
component. The non-thermal component consists of a power-law with 
spectral index $\gamma$ and flux $F_{50}$ at normalization energy 50 keV,
a low energy turnover at energy  $E_\mathrm{turn}$ and
a high-energy break at energy $E_\mathrm{br}$. The power-law index above
$E_\mathrm{br}$ is named $\beta$. The index below $E_\mathrm{turn}$ is fixed at 1.5.
The model is illustrated in Fig.~\ref{f_multi}. 
It has 9 free parameters, but it was not always necessary to let all
of them free.
At first, an isothermal component ($T_1$, $EM_1$) was fitted, followed by a
power-law with a low-energy turnover (henceforth called 'standard model'). If the
non-thermal spectrum showed indications of a break at high energies, the
parameters $E_\mathrm{br}$ and $\beta$ were also fitted. Some of the larger flares
showed strong thermal emission that was fitted with an additional hotter isothermal component
($T_2$, $EM_2$).

\setlength{\tabcolsep}{1.3mm}
\begin{sidewaystable*}
\begin{minipage}[t][14cm]{\columnwidth}

\caption{Event list}
\label{eventlist}
\renewcommand{\footnoterule}{}
\begin{tabular}{lllllllrllllllrllllll}   
\hline \hline 
 
  &          &       &                                                  &&&&&&&&&&&&&&&&\\
No.  & Date     & Time   & GOES\footnote{Background subtracted GOES-class}
&GOES\footnote{Original GOES-class before background subtraction}   & $\gamma$
&$F_{35}$\footnote{units: photons~s$^{-1}$cm$^{-2}$keV$^{-1}$} &No.&  Date  & Time &
GOES$^{a}$&GOES$^{b}$ & $\gamma$ &$F_{35}$$^{c}$ & No.&Date&Time&GOES$^{a}$&GOES$^{b}$ & $\gamma$&$F_{35}$$^{c}$ \\       
&&&&orig.&&&&&&&orig.&&&&&&&orig.&&\\

\hline					         
  1 &22-Aug-02&01:52&M5.7&M5.9&3.4&1.1$\cdot$10$^{1}$ &32&18-Sep-02&01:12&C2.9&C4.0&5.6&2.2$\cdot$10$^{-2}$&63&11-Apr-03&10:27&B2.0&B3.5& 5.1&5.5$\cdot$10$^{-3}$   \\ 
  2 &10-Jun-03&11:06&M5.2&M5.4&2.9&1.7$\cdot$10$^{0}$ &33&29-Aug-02&13:22&C2.8&C5.0&4.6&3.5$\cdot$10$^{-2}$&64&23-Nov-02&02:45&B1.7&B9.2& 5.3&2.2$\cdot$10$^{-2}$  \\
  3 &20-May-02&10:52&M5.1&M5.3&3.6&5.9$\cdot$10$^{-1}$&34&20-Jul-03&14:51&C2.6&C3.8&8.3&1.4$\cdot$10$^{-3}$&65&12-May-03&17:38&B1.3&B2.4& 5.0&9.7$\cdot$10$^{-4}$  \\
  4 &20-Feb-02&09:58&M4.8&M4.9&3.6&6.9$\cdot$10$^{-1}$&35&30-Aug-02&04:29&C2.2&C3.5&4.7&2.0$\cdot$10$^{-1}$&66&27-Sep-03&10:30&B1.2&B4.8& 4.2&6.4$\cdot$10$^{-3}$   \\
  5 &15-Apr-02&00:10&M4.0&M4.1&5.8&6.8$\cdot$10$^{-1}$&36&28-Feb-02&12:07&C2.0&C4.5&5.5&3.6$\cdot$10$^{-2}$&67&02-Dec-02&12:44&B1.1&B5.1& 6.7&3.3$\cdot$10$^{-4}$ \\
  6 &18-Nov-03&08:08&M3.8&M3.9&2.6&2.3$\cdot$10$^{0}$ &37&29-Nov-03&18:01&C2.0&C2.7&7.5&6.8$\cdot$10$^{-3}$&68&10-Jan-04&01:04&B1.1&B4.1& 7.9&9.1$\cdot$10$^{-5}$  \\
  7 &29-Sep-02&06:36&M3.0&M3.1&3.6&3.5$\cdot$10$^{0}$ &38&15-Nov-02&13:47&C1.7&C3.4&6.4&2.6$\cdot$10$^{-2}$&69&23-Mar-04&13:12&B1.0&B3.7& 5.2&1.1$\cdot$10$^{-3}$  \\
  8 &13-Jun-03&02:01&M2.8&M3.1&6.9&7.0$\cdot$10$^{-1}$&39&02-Oct-02&21:15&C1.4&C2.1&6.6&9.5$\cdot$10$^{-3}$&70&14-Apr-03&12:19&A9.7&B2.4& 4.4&8.9$\cdot$10$^{-3}$  \\
  9 &16-Apr-02&13:10&M2.7&M2.8&6.5&3.2$\cdot$10$^{-1}$&40&09-Sep-02&21:31&C1.3&C2.2&2.4&5.0$\cdot$10$^{-2}$&71&24-Feb-03&20:39&A9.3&B1.9& 4.9&3.3$\cdot$10$^{-3}$  \\
 10 &06-Jul-02&03:32&M2.0&M2.1&5.8&3.6$\cdot$10$^{-1}$&41&05-Apr-03&00:34&C1.3&C2.1&5.1&1.4$\cdot$10$^{-2}$&72&25-Feb-03&01:43&A9.3&B1.9& 6.8&1.8$\cdot$10$^{-4}$ \\
 11 &27-Sep-02&13:08&M1.9&M2.0&3.3&2.4$\cdot$10$^{-1}$&42&24-Feb-02&15:37&C1.1&C2.5&5.1&1.1$\cdot$10$^{-1}$&73&24-Feb-03&17:18&A8.5&B1.9& 4.7&2.2$\cdot$10$^{-3}$  \\
 12 &20-Oct-02&00:41&M1.8&M1.9&5.3&3.9$\cdot$10$^{-1}$&43&28-May-02&03:12&B8.8&C3.4&8.2&2.6$\cdot$10$^{-3}$&74&10-Mar-03&13:31&A7.2&B4.2& 3.3&8.9$\cdot$10$^{-3}$  \\
 13 &10-Apr-02&19:02&M1.7&M1.8&2.9&3.8$\cdot$10$^{0}$ &44&27-Nov-02&09:05&B8.7&C1.7&7.3&9.6$\cdot$10$^{-3}$&75&30-Mar-03&03:50&A6.9&B6.0& 4.3&1.1$\cdot$10$^{-2}$  \\
 14 &01-Jun-02&03:53&M1.5&M1.6&2.3&2.8$\cdot$10$^{0}$ &45&01-Dec-03&15:02&B7.7&C1.4&6.8&4.7$\cdot$10$^{-3}$&76&12-Apr-03&04:14&A5.6&B1.8& 3.0&3.7$\cdot$10$^{-2}$   \\
 15 &21-Aug-02&01:39&M1.4&M1.6&2.7&1.3$\cdot$10$^{1}$ &46&01-Jul-02&04:44&B5.8&C1.2&5.5&6.0$\cdot$10$^{-3}$&77&14-Apr-03&06:04&A5.0&B1.7& 5.1&2.0$\cdot$10$^{-3}$  \\
 16 &04-Apr-02&10:44&M1.3&M1.5&5.2&2.6$\cdot$10$^{-1}$&47&28-Mar-02&05:06&B5.4&C1.4&3.3&2.3$\cdot$10$^{-1}$&78&24-Dec-03&11:01&A4.6&B4.2& 5.7&1.8$\cdot$10$^{-3}$  \\
 17 &15-Apr-02&23:10&M1.2&M1.3&4.9&6.6$\cdot$10$^{-1}$&48&23-Sep-03&11:42&B4.4&B8.1&6.7&2.8$\cdot$10$^{-4}$&79&14-Apr-03&23:00&A4.4&B1.9& 3.9&3.4$\cdot$10$^{-3}$  \\
 18 &21-Feb-02&18:10&C9.9&M1.2&5.3&4.5$\cdot$10$^{-1}$&49&28-Nov-02&04:36&B3.8&C1.1&2.9&7.7$\cdot$10$^{-1}$&80&26-May-03&12:25&A4.3&B3.8& 5.4&2.8$\cdot$10$^{-3}$  \\
 19 &17-Apr-02&00:38&C9.9&M1.1&4.3&2.5$\cdot$10$^{-1}$&50&11-Aug-02&21:48&B3.6&C1.5&5.0&6.3$\cdot$10$^{-2}$&81&17-Apr-03&13:36&A4.2&B2.9& 5.6&1.2$\cdot$10$^{-3}$  \\
 20 &20-Feb-02&16:23&C9.3&M1.1&3.9&1.6$\cdot$10$^{-1}$&51&19-Jul-02&08:38&B3.3&C1.0&2.9&6.3$\cdot$10$^{-2}$&82&30-Jul-03&08:38&A3.8&B2.0& 4.0&3.7$\cdot$10$^{-3}$  \\
 21 &23-Aug-02&11:58&C8.4&M1.4&6.8&1.9$\cdot$10$^{-1}$&52&03-Aug-02&22:22&B3.0&C2.0&6.5&1.5$\cdot$10$^{-2}$&83&14-Mar-04&10:09&A2.8&B1.6& 3.9&4.4$\cdot$10$^{-3}$  \\
 22 &18-Jul-02&23:14&C7.9&C8.6&4.0&1.0$\cdot$10$^{0}$ &53&22-Apr-03&01:27&B3.0&B7.4&4.3&2.3$\cdot$10$^{-3}$&84&12-Apr-03&08:50&A2.8&B1.8& 2.3&2.8$\cdot$10$^{-2}$  \\
 23 &30-Aug-02&02:40&C7.7&C8.9&4.2&5.4$\cdot$10$^{-1}$&54&28-Jul-03&03:25&B2.9&B5.3&4.4&4.9$\cdot$10$^{-3}$&85&10-Apr-03&13:42&A2.5&B1.7& 3.2&1.1$\cdot$10$^{-2}$  \\
 24 &24-Oct-02&18:05&C7.4&C7.8&3.0&5.6$\cdot$10$^{-1}$&55&11-Jan-04&16:09&B2.7&B6.2&6.1&8.4$\cdot$10$^{-4}$&&         &    &     &      &  &             \\
 25 &20-Aug-02&22:28&C6.8&C8.4&3.5&1.9$\cdot$10$^{0}$ &56&25-Oct-02&16:21&B2.5&B7.8&5.8&2.6$\cdot$10$^{-4}$&86&23-Sep-02&04:48&C1.3&C1.6&   &  \\
 26 &29-Aug-02&23:32&C6.2&C7.6&5.8&1.9$\cdot$10$^{-1}$&57&26-Nov-03&22:40&B2.5&B8.7&6.5&6.9$\cdot$10$^{-4}$&87&09-Sep-02&16:32&B8.0&C2.2&   & \\
 27 &16-Sep-02&20:02&C4.6&C5.3&3.8&2.1$\cdot$10$^{-1}$&58&15-Dec-03&11:36&B2.5&B6.8&6.2&4.6$\cdot$10$^{-3}$&88&04-May-02&10:21&B4.4&C1.3&   &  \\
 28 &09-Apr-02&06:05&C4.2&C7.8&7.5&2.0$\cdot$10$^{-2}$&59&08-Mar-03&11:54&B2.3&B7.0&3.9&6.3$\cdot$10$^{-3}$&89&17-Nov-02&07:18&B4.0&C1.7&   &  \\
 29 &02-Apr-02&02:28&C3.9&C5.6&6.8&2.4$\cdot$10$^{-2}$&60&18-Mar-04&00:23&B2.3&B5.5&5.1&5.3$\cdot$10$^{-3}$&90&01-Mar-04&02:10&A1.4&B1.7&   &  \\
 30 &23-Aug-02&00:43&C3.6&C5.4&5.4&1.5$\cdot$10$^{-1}$&61&05-Mar-04&18:25&B2.2&B8.0&5.0&1.7$\cdot$10$^{-2}$&91&29-Feb-04&20:09&A5.6&B2.5&   &  \\
 31 &31-Aug-02&14:22&C3.2&C5.1&2.9&9.8$\cdot$10$^{-1}$&62&11-Apr-03&12:02&B2.0&B3.7&4.8&8.4$\cdot$10$^{-3}$&92&28-May-02&02:00&B5.2&C2.1&   &  \\

\hline
\end{tabular}

\end{minipage}
\end{sidewaystable*}

All 92 flares are
listed in Table~\ref{eventlist} with a number, date, peak time, GOES flux,
spectral index and non-thermal flux.
In total, 6 different models where fitted: 
\begin{enumerate}
\item One thermal part and an unbroken power-law (standard model): This
  provided the best fitting for 31 
  flares. Another 37 flares where fitted with 
this model, but were also consistent with two thermal components and no power-law. This point
  will be discussed later.  
\item One thermal part and a broken power-law:
For 15 of the larger flares, a broken power-law fitting was applied. Most of these
flares had emission beyond 50~keV, some up to 100~keV or more. Three of them
can also be fitted with two thermal components. 
\item Two thermal parts plus power-law:
2 flares had thermal emission up to around 20~keV which we fitted with
two temperatures instead of one. 
\item Only thermal: 6 flares show no indications for a
power-law part in the spectrum 
and were therefore fitted with one thermal part only (flare numbers 86--90
in Table~\ref{eventlist}) and with only two thermal components (no. 91). The largest of them was
a C1.3 after background subtraction. These
flares were not used in the further analysis. 
\item Only broken power-law:
Flare no. 92 (Table~\ref{eventlist}) was particularly difficult to fit. The best fitting was a
broken power-law only. The spectrogram shows a very impulsive, short emission,
even at the lowest energies. If present at all, the thermal emission must have
been very small. This flare was not taken into the further analysis.
\end{enumerate} 
In conclusion, 85 out of 92 events could be 
fitted with the standard model or additionally with a  break or a second 
isothermal component.

\subsection{Flares with more than one fitting model} \label{2model}
For many of the flares smaller than background subtracted GOES class C2,
fitting a model with only 2
thermal components and no non-thermal emission was also possible.
These smaller flares are less intensive than large flares. Therefore the
non-thermal emission can be quite weak
or even lie below the background level. As the signal-to-noise ratio decreases,
different fittings are possible, but not necessarily plausible.
The $\chi^2$ of the two different models are often about the same value,
making it impossible to determine which model is better. The temperatures
range up to 260 MK in B-class 
flares. Although the thermal energy content would not be unphysically high, the standard model
has been taken for each of those flares in the further analysis, as the
analogy to large flares is the simplest assumption. 


\section{Results} \label{results}
We present here the results obtained from the fittings, comparing several
flare properties to each other.
\subsection{Non-thermal emission}

\subsubsection{Spectral index versus non-thermal photon flux}
The non-thermal photon flux is modeled by a power-law
distribution with spectral index $\gamma$. For events with a broken power-law,
$\gamma$ refers to the spectral index below the break energy.
For the comparison of the spectral index and the photon flux we cannot use the
total flux above $E_\mathrm{turn}$,
 $F_\mathrm{tot}=\int_{E_\mathrm{turn}}^\infty F(E)\, \mathrm{d}E$, since
this quantity depends strongly on $E_\mathrm{turn}$ which is poorly determined
by the observations. 
Instead we use the flux at a reference energy
$E_0$. The maximum energy with still observable flare emission varies from
about 20~keV to 300~keV, and increases with flare
size. Therefore, choosing a high $E_0$ means extrapolating the flux of small 
flares into a range where it is not actually observed, whereas choosing a
low $E_0$ means extrapolating the flux of large flares into a range where thermal
emission might dominate. 
The influence of the choice of $E_0$ on the
plot of 
$\gamma$ vs. $F_{E_0}$ is described in GB04.  
We chose $E_0$=35~keV as a typical energy for the presentation of the
results. This allows also for comparisons with previous work. 

\begin{figure} 
\resizebox{\hsize}{!}{\includegraphics{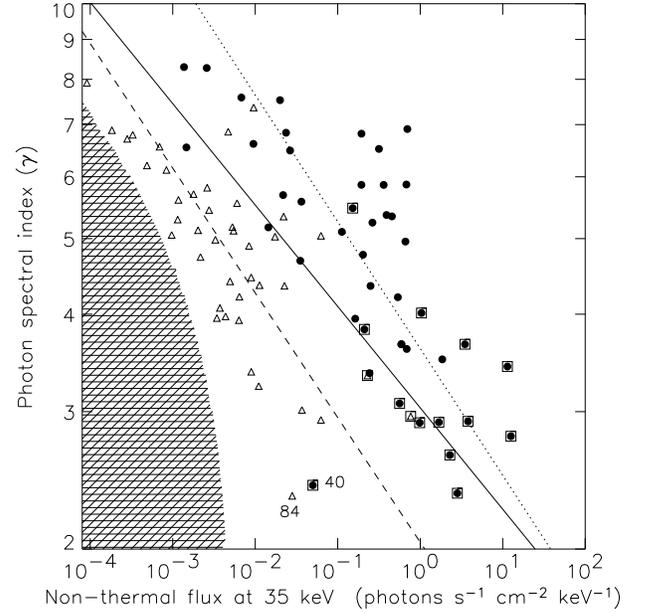}}
\caption {\small Spectral index $\gamma$ versus RHESSI non-thermal
  photon flux at 35~keV, $F_{35}$. The 
  solid line  represents the linear bisector regression, given in
  Eq.(\ref{f35vsgamma}). Flares with GOES class
  smaller than C2 are indicated with a triangle, those larger than C2
  with a circle.  The points outlined by a
  square are the events which where fitted with a broken power-law. The dotted
  line represents the regression for events larger 
  than C2, and the dashed line represents the regression for events smaller
  than C2. Events in the shaded region would not appear in the flare list and
  would not be selected. The numbers
  refer to the event list in Table~\ref{eventlist}. These events are
  discussed in section \ref{exceptional}.}
\label{fluxvsgamma}
\end{figure}

The plot of $\gamma$ vs. $F_{35}$ for the examined 85 flares with a non-thermal
component is shown in Fig.~\ref{fluxvsgamma}. 
Despite the large scatter of the data, a linear correlation in the double
logarithmic plot
is noticeable. A linear regression has been performed. As suggested by Isobe
et al. (\cite{Isobe90}) for this kind of scattered data, the bisector regression method
has been used. 
The relation can be written as a power-law model
\begin{equation} \label{powerlaw}
\ln \gamma=\ln(A)-\alpha\ln F_{35},
\end{equation}
with statistical errors $\ln(A)=1.11\pm0.06$ and $\alpha=0.13\pm0.01$, or 
\begin{equation} \label{f35vsgamma}
 \gamma=3.0F_{35}^{-0.13 }.
\end{equation}
Since the data have a fairly large scatter and there are less than 100
data points, it is difficult to determine
unambiguously a single best regression model. The ordinary least squares
regressions of y vs. x  and x vs. y yield a limit for the confidence 
range of $\alpha=0.13\pm 0.07$.

As discussed in Sec.~\ref{2model} there was more than one possible fitting
model for most of the events smaller than C2 (after background subtraction). 
Therefore, the set has been divided into
events smaller and larger than
C2, and an independent regression has been made for both parts. The non-thermal
flux of both sets ranges over 4 orders of magnitude and their spectral
index ranges from 2.3 up to 8.3. The regression of the low GOES-class flares leads to a relation
\begin{equation} \label{lowclass}
 \gamma=2.04F_{35}^{-0.16}, 
\end{equation}
with confidence range 0.16$\pm0.05$ for the exponent.
These flares are indicated with a triangle in Fig.~\ref{fluxvsgamma}.
When taking only the events larger than C2, the relation becomes
 \begin{equation} \label{highclass}
 \gamma=3.60F_{35}^{-0.16},
\end{equation}
with confidence range 0.16$\pm$0.06 for the exponent. 
Both sets of flares thus yield the same slope in logarithmic $F_{35}$ vs. $\gamma$
dependence. For the flares larger than C2 however, the regression line is shifted to
higher $F_{35}$ values by about one order of magnitude (compare
Fig.~\ref{fluxvsgamma}). 
This leads to a flatter
slope, when the regression is performed for the whole data set. 

\begin{figure} 
\resizebox{\hsize}{!}{\includegraphics{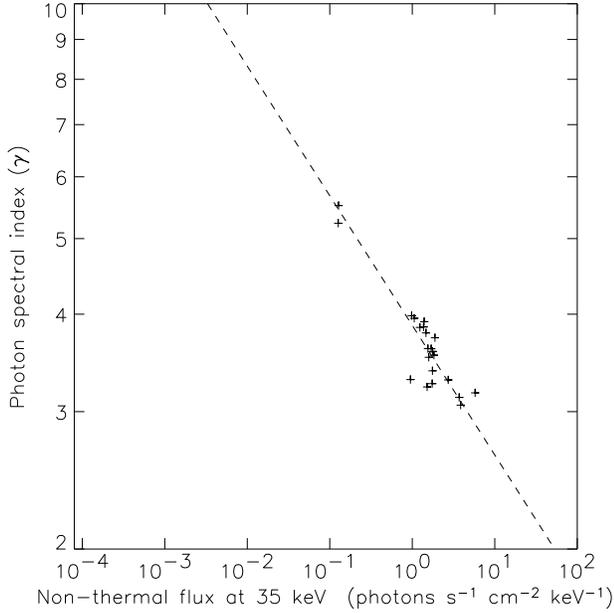}}
\caption {\small Spectral index ($\gamma$)
  versus non-thermal photon flux at 35 keV for several sub-peaks of a single
  flare. The dashed line represents the regression
  line (Eq. \ref{paolosflareeq}).}
\label{paolosflare}
\end{figure}

We are interested in the difference between small and large flares and whether
their non-thermal parameters have something in common with the parameters of
several sub-peaks of one single flare.  
Therefore, the same plot and regression has been made for sub-peaks
of one of the flares that have been analyzed by GB04 (event no. 23 in their
list, not present in our list). This plot is
shown in Fig.~\ref{paolosflare}. The relation is
\begin{equation} \label{paolosflareeq}
 \gamma=3.89F_{35}^{-0.17},
\end{equation}
and the confidence range for the exponent $\alpha=0.17\pm 0.01$.
The relation between $\gamma$ and $F_{35}$ is quite similar for several peaks
of one flare as for many flares of different size at time of maximum
non-thermal emission.  
Note that the maximum non-thermal sub-peak flux of the inspected
flare (GOES M4.9) ranges from 10$^{-1}$ to 10$^1$~photons~s$^{-1}$cm$^{-2}$keV$^{-1}$
 and therefore represents only a small 
section of the flux range of all flares in the presented sample. The observed
power-law component (Eq. \ref{paolosflareeq}) is, however, very close to the
relation for only events with GOES class larger than C2.

\subsubsection{Observational limits}
As mentioned in Sect.~\ref{selectioneffect} there is a selection effect
due to the nature of the flare list (only events
with emission above background in the 12--25 keV band are listed). 
We made an estimate on how the $\gamma$ vs. $F_{35}$ relation is influenced by this
selection effect. 

The idea is to simulate count spectra and find limits for $F_{35}$ and $\gamma$
for which the total number of counts in the 12--25 keV energy range is larger
than 3$\sigma$ of the background.
The spectral response matrix (SRM) without attenuators has been used.
The average background in counts s$^{-1}$ in the
12--25~keV energy band was estimated from the light curve.
Photon spectra have been calculated, assuming a thermal component and an unbroken power-law. The
temperature and emission measure were assumed according to the relations 
given in Eqs.~(\ref{tvsflux35eq}) and
(\ref{emvsflux35eq}). The low
energy turnover was fixed at 9~keV. The photon spectra 
have been calculated for several values
of spectral index $\gamma$ from the interval [2,10] for a
number of non-thermal fluxes $F_{35}$ from the interval
$[10^{-5},10^{-2}]$ photons s$^{-1}$ cm$^{-2}$ keV$^{-1}$. Each of these
photon spectra has been multiplied by the SRM to get count spectra. For
each flux $F_{35}$ and spectral index $\gamma$ in the above given intervals,
the total number of counts in the energy range 12--25 keV has been calculated
and compared to the background counts. For every $F_{35}$ there exists a
limit in $\gamma$ where the total simulated counts are less than 3$\sigma$
of the background counts. These boundaries plotted in the $\gamma$ vs. $F_{35}$
relation define a region (shaded region in Fig.~\ref{fluxvsgamma}) where
events do not appear in the flare list.

From the location of the observational limits and the data points on the plot,
one can expect the selection effect to be largest for soft events with a 
small non-thermal flux. If such events are missing in our selection, the
regression line would be flatter in reality.

\subsubsection{Energy of 'equal photon flux'?}
As weaker flares are softer, there must be a region where spectra of weak and
strong flares intersect. A 'pivot region' is sometimes found for the
SHS evolution of single subpeaks (see Introduction). Similarly, an energy
E$_\mathrm{eq}$ or an energy range may exist where small and large flares have
equal non-thermal photon flux F$_\mathrm{eq}$. 

We have searched for an 'energy of equal photon flux' by calculating the energy distribution of
all intersections between the non-thermal power-law fittings. The half width is
extremely broad, ranging from 2.1 to 52 keV with a peak at 10.4~keV.

\subsubsection{High-energy break}
As mentioned before, 15 spectra have been fitted with a high-energy
break.  These events are indicated with a square in
Fig.~\ref{fluxvsgamma}. The non-thermal
flux $F_{35}$ of all these events is higher than $4\cdot10^{-2}$ photons
s$^{-1}$cm$^{-2}$keV$^{-1}$. Thus high-energy breaks are observed only in
large flares. 

The break energy does not correlate with the spectral index. 
However a correlation appears between the spectral index
below break energy ($\gamma$) and the spectral index above break energy
($\beta$) following a relation $\beta=(1.96\pm0.20)\gamma-1.59$.
A similar relation has previously been reported by Lin
\& Schwartz (\cite{Lin86}). Further, we observed a
correlation between the break energy and the non-thermal photon flux at 35
keV, indicating that flares with smaller non-thermal flux have on the average a lower
break energy. Such a comparison has previously been made by Dulk et al. (\cite{Dulk92}),
who found no correlation between break energy and non-thermal flux.

However, we were not able to exclude
without doubt selection effects that could lead to these correlations (see
also discussion in Sect.~\ref{discussion}).

\subsection{Thermal flare properties compared to non-thermal properties}
In a next step the soft X-ray flux observed by the GOES satellites was
included in the analysis. The GOES flux of all flares at time of maximum 
emission (GOES peak time) in the 1--8~\AA\ band has been extracted. For a
proper analysis, the background was subtracted. The background was
chosen in
GOES light curves as either linearly
interpolated between pre-flare and post-flare background or as constant. 
Subtracting background changes the GOES class contrary to the usual
classification. When referring to the size of a flare, the GOES class
according to peak flux without background subtraction is given throughout this
work. For comparing physical parameters,
however, the background-subtracted values have been used. 

\begin{figure}
\resizebox{\hsize}{!}{\includegraphics{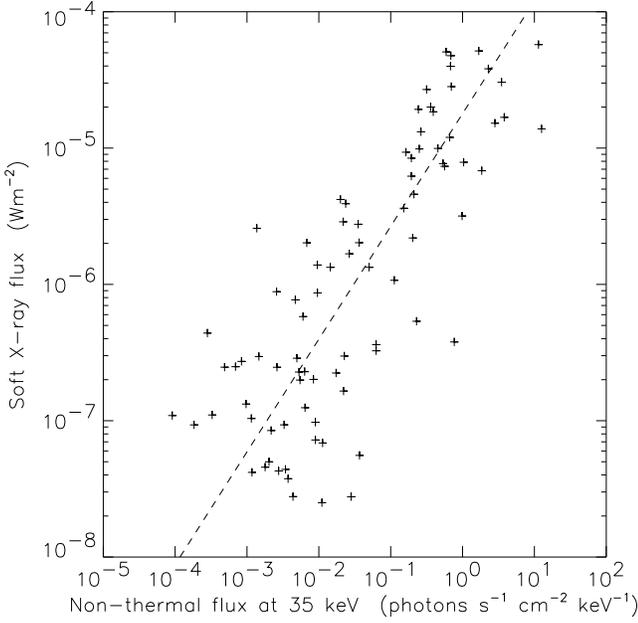}}
\caption{\small Soft X-ray flux $F_\mathrm{G}$ versus non-thermal flux $F_{35}$ at 35
  keV  with regression line, as given in Eq.~(\ref{f35vsfgoeseq}).}
\label{f35vsfgoes}
\end{figure}

First, the non-thermal flux $F_{35}$ has been compared to the maximum GOES
flux, $F_\mathrm{G}$. There
is a linear correlation in a double logarithmic plot, presented in Fig.~\ref{f35vsfgoes}, giving:
\begin{equation} \label{f35vsfgoeseq}
F_\mathrm{G}=1.8\cdot 10^{-5}F_{35}^{0.83},
\end{equation}
where the exponent $b=0.83 \pm 0.19$. $F_\mathrm{G}$ is in units of Wm$^{-2}$ and $F_{35}$
in photons $\mathrm{s^{-1}cm^{-2}keV^{-1}}$. 

\begin{figure} 
\resizebox{\hsize}{!}{\includegraphics{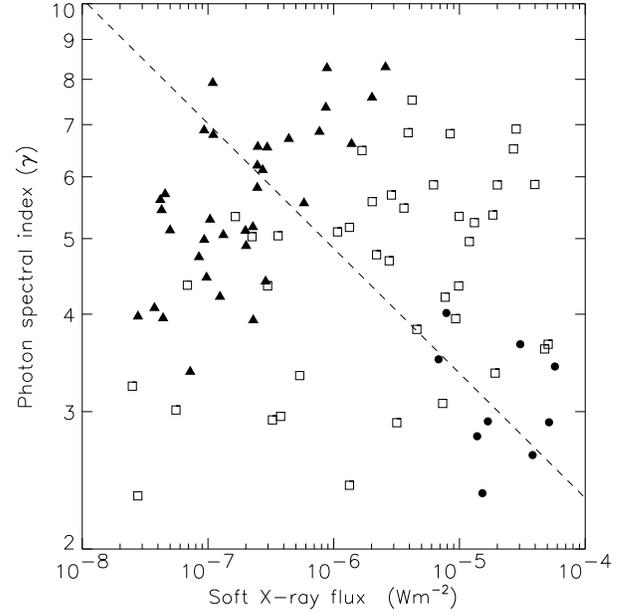}}
\caption {\small Spectral index $\gamma$ versus soft X-ray flux
  $F_\mathrm{G}$. The data points  have been labeled with 3 different symbols according to
 3 flux intervals in $F_{35}$ (units: photons $\mathrm{s^{-1}cm^{-2}keV^{-1}}$).
  Triangles: $F_{35}$
  in the interval [10$^{-5}$,10$^{-2}$]; squares: $F_{35}$
  in the interval [10$^{-2}$,10$^0$]; circles: $F_{35}$
  in the interval [10$^0$,10$^2$]. The dashed line is the expected trend-line
  from the correlations of $\gamma$ vs. $F_{35}$, and $F_\mathrm{G}$ vs. $F_{35}$.}
\label{gammavsgoesfluxsymb}
\end{figure}

As $F_{35}$ vs. $\gamma$ and $F_{35}$ vs. $F_\mathrm{G}$ correlate, one must expect
a correlation between $F_\mathrm{G}$ and $\gamma$. 
Figure~\ref{gammavsgoesfluxsymb} illustrates that this is not the case.  From the relations 
  $\gamma=AF_{35}^{\alpha}$ and $F_\mathrm{G}=CF_{35}^d$ one can calculate an expected
  relation $\gamma=AC^{-\alpha/d}F_\mathrm{G}^{\alpha/d}$, giving
  $\gamma=0.53F_\mathrm{G}^{-0.16}$, represented by the dashed line.
As expected, the flares with small $F_{35}$ are in the upper left corner and the
ones with large $F_{35}$ in the bottom right. However, the scatter in $\gamma$
is so large that the correlation is statistically not significant. 

\begin{figure}
\resizebox{\hsize}{!}{\includegraphics{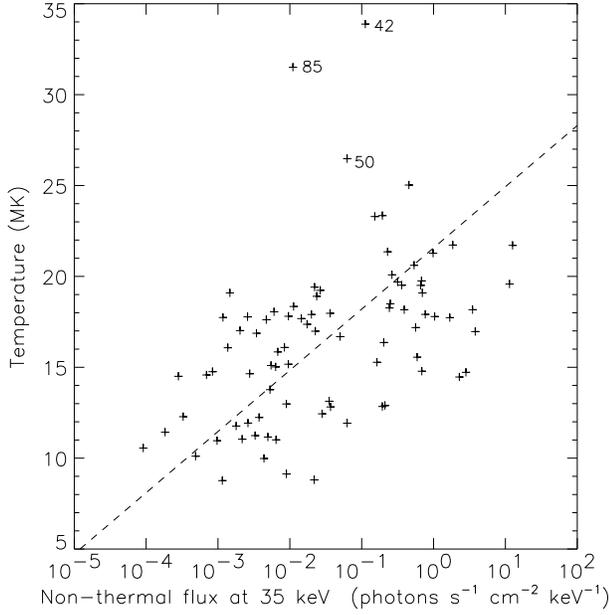}}
\caption {\small RHESSI temperature $T_1$ versus non-thermal flux at 35~keV
  with linear bisector regression line (Eq.~\ref{tvsflux35eq}). The numbers
  refer to the event list in Table~\ref{eventlist}.}
\label{tvsflux35}
\end{figure}

\begin{figure} 
\resizebox{\hsize}{!}{\includegraphics{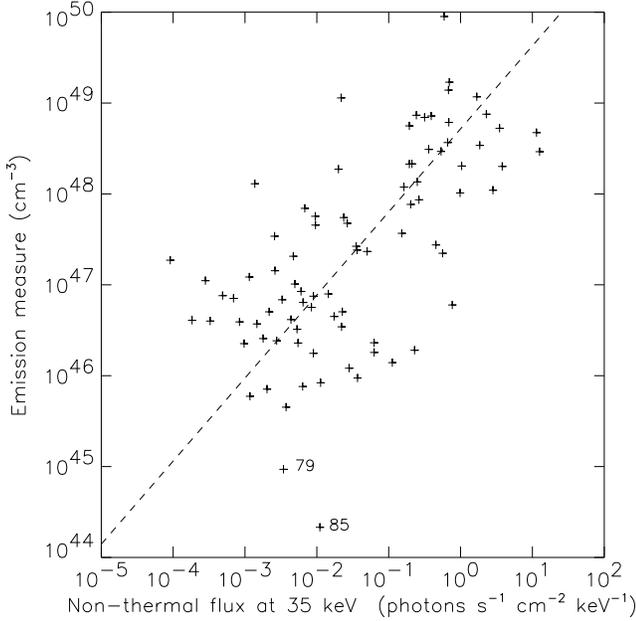}}
\caption {\small Emission measure $EM_1$ versus non-thermal flux at
  35~keV. The dashed line indicates the linear bisector regression line
  (Eq.~\ref{emvsflux35eq}). The
  numbers refer to the event list in Table~\ref{eventlist}.}
\label{emvsflux35}
\end{figure}

We compared also the temperature $T_1$ and emission measure $EM_1$ of the thermal
plasma derived from RHESSI data, to the non-thermal X-ray flux at 35 keV.
Figure~\ref{tvsflux35} indicates
a correlation between temperature and non-thermal flux with a relation
\begin{equation} \label{tvsflux35eq}
T_1=1.46\ln F_{35}+21.57,
\end{equation}
where $T_1$ is in units of 10$^6$ K and $F_{35}$ in
photons~$\mathrm{s^{-1}cm^{-2}keV^{-1}}$. 
A higher non-thermal flux is therefore associated with a higher temperature.
A similar observation holds for the relation between $F_{35}$ and the emission
measure (Fig.~\ref{emvsflux35}). 
\begin{equation} \label{emvsflux35eq}
EM_1=5\cdot 10^{48}F_{35}^{0.91},
\end{equation}
with $EM_1$ in $\mathrm{cm^{-3}}$.

\subsection{Thermal flare plasma}
\begin{figure} 
\resizebox{\hsize}{!}{\includegraphics{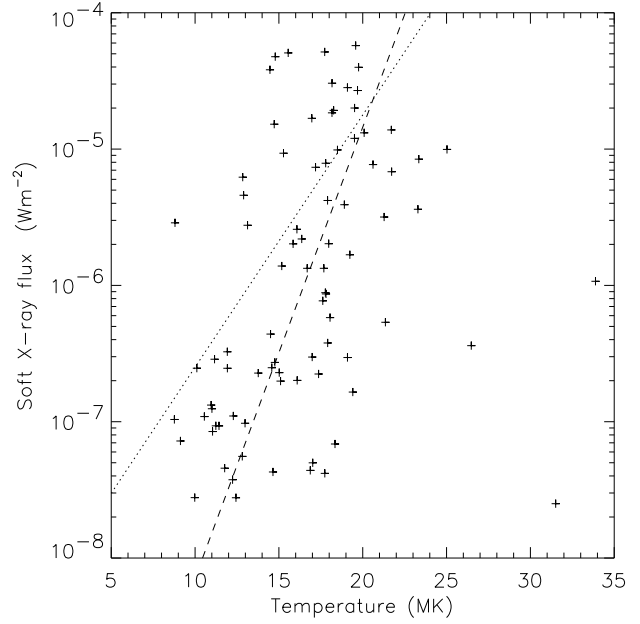}}
\caption {\small Maximum soft X-ray flux versus temperature. The dashed line
  gives the linear bisector regression to the data (Eq. \ref{tvsgoesfluxeq}). The dotted line is
  the trend line calculated from the Feldman et al. result (Eq. \ref{feldmaneq}).}
\label{tvsgoesflux}
\end{figure}

Similar to the investigations by Feldman et al. (\cite{Feldman96}), we also compared the
temperature $T_1$ at hard X-ray peak time to the maximum soft X-ray flux
$F_\mathrm{G}$. We note a correlation in
Fig.~\ref{tvsgoesflux}, giving 
\begin{equation} \label{tvsgoesfluxeq}
 F_\mathrm{G}=3.5\cdot 10^{0.33T_1-12},
\end{equation}
where $F_\mathrm{G}$ is in $\mathrm{Wm^{-2}}$, plotted in a logarithmic
scale, and $T_1$ is in units of 10$^6$~K, plotted in linear scale. Due to
the large scatter, the power factor p in Eq.~(\ref{tvsgoesfluxeq}) has a
confidence range of p=$0.33\pm0.29$.
The Feldman et al. result ($F(T)=3.5\cdot 10^{0.185 T - 9.0}$) differs considerably, as is illustrated in
Fig.~\ref{tvsgoesflux}. It is though still within the estimated range for the
regression parameters.
Contrary to Feldman et al., we used background subtracted GOES flux.  The
background-subtraction from the GOES data has the biggest influence on the smaller flares,
altering the flux to smaller values and extending the flux range one order
of magnitude. Using the original, not background
subtracted GOES data, we get $F_\mathrm{G}=1.7\cdot 
10^{0.25T_1-10}$.

\subsubsection{Comparison of temperatures} \label{comparetemp}
\begin{figure} 
\resizebox{\hsize}{!}{\includegraphics{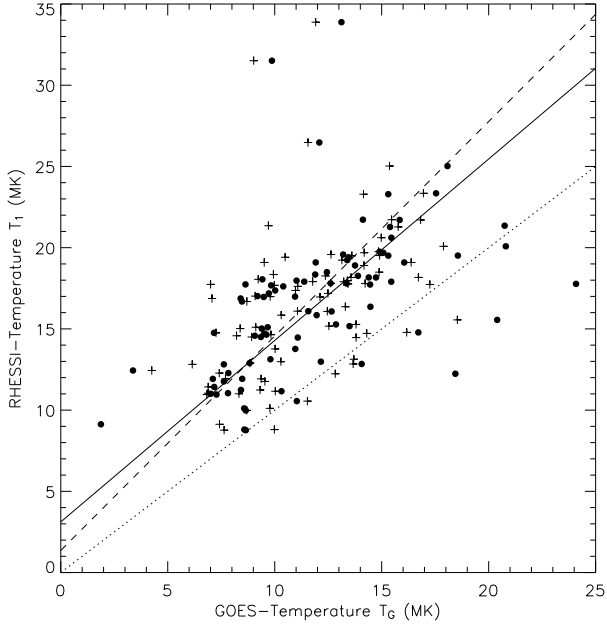}}
\caption{\small Temperature measured by
  RHESSI ($T_1$) compared to GOES temperature ($T_\mathrm{G}$). The dotted
  line indicates the diagonal line. Crosses indicate the GOES temperature at
  GOES peak time, the dashed line gives the corresponding regression
  line. Dots give the GOES temperature measured at the same time as the RHESSI
  temperature with the solid line as regression line.}
\label{comparetemperatures}
\end{figure}

From the flux ratio in the 2 GOES wavelength bands, one can calculate an
isothermal temperature $T_\mathrm{G}$ at the time of maximum soft X-ray flux, using
Mewe spectral models (Mewe et al. \cite{Mewe85}).
As the minimum energies of the two bands
are 1.5~keV and 3.1~keV respectively, GOES measurements are dominated by
photons at lower energy than used to determine the temperature from RHESSI
data ($\ge$4 keV). A comparison of the two measurements is also a check on the applied
methods: flux ratio (GOES, at soft X-ray peak time) versus spectral fitting
(RHESSI, at hard X-ray peak time).
Figure~\ref{comparetemperatures}  
indicates that the temperature derived from RHESSI data is often
higher. A linear bisector regression yields a relation
$T_1=1.31T_\mathrm{G}+1.47$ (in MK), but is influenced by the three
outliers. Without the outliers, the relation would be $T_1=1.12T_\mathrm{G}+3.12$.

The emission measure derived from the GOES data is higher than that from the
RHESSI data for 82 out of 85 events. The ratio $EM_1/EM_\mathrm{G}$ increases 
for larger events. On the
average, $EM_1\approx 0.1EM_\mathrm{G}$. Most likely, the
RHESSI measurements emphasize the hotter part of a non-isothermal temperature
distribution in the soft X-ray source.
Part of these trends could also be accounted for by the different times of
measurements, as the hard X-ray peak usually occurs earlier in the course of
a flare than the soft X-ray peak. Taking the GOES temperature at the
  same time as the RHESSI temperature leads indeed to a relation with less
  pronounced trend: $T_1=1.13T_\mathrm{G}+3.17$, but the RHESSI temperature is
  still higher than the GOES temperature in most cases (Fig. \ref{comparetemperatures}).

\subsection{Inspection of exceptional flares} \label{exceptional}
In most of the presented plots there are a few events far from the overall trend line.
The most striking ones have been numbered according to Table~\ref{eventlist}
and are shortly discussed here.  \\ \\
\textbf{No. 40 and 84} have very hard spectral indices, no. 84
has the hardest $\gamma$ in our data set. Its spectrogram shows short,
impulsive emission up to over 30 keV. \\ \\
\textbf{No. 42, 50 and 85} have very high temperatures. 
 The spectrograms of all three events
indicate a very short, impulsive, non-thermal peak with emission over 30~keV. \\ \\
\textbf{No. 79} has a small emission measure of only $9.3\cdot
10^{44}$~cm$^{-3}$. 
The temperature of 16.9 MK is rather high for a flare of this
size. Again, the flare is very short and impulsive. \\ \\ 

The common characteristic of all these flares is the short duration of the
non-thermal emission. 

\section{Discussion}\label{discussion}
The high spectral resolution of RHESSI's germanium detectors permits the
 separation of a non-thermal and one or two thermal components in most X-ray
 spectra of solar flares over a large range of flare sizes. The selection was
 made according to the usual definition of flare size in terms of peak soft
 X-ray flux (labeled B1 to M6) as measured by the GOES 1 - 8 \AA\ channel. It
 remains nearly uniform in soft X-ray peak flux after background subtraction
 but extends one more order of magnitude to smaller flares (A1 to M6). 

The analysis of the non-thermal emission shows a clear relation between the
 spectral index $\gamma$ and the power-law normalization $F_{35}$ at peak
 time. The correlation between $\gamma$ and $F_{35}$ can be expressed by a
 power-law. We note that the quantitative relation is not
 identical to the relation found for the SHS behavior in subpeaks of
 events. The spectral index $\gamma$ depends less on $F_{35}$, and the
 power-law exponent $\alpha$ is smaller (Eq.~\ref{f35vsgamma}). The two
 exponents $\alpha$ for the
 temporal evolution of single flares and the peak values of many flares can be
 brought closer together when we accept that the flares smaller in soft X-rays
 (and GOES class) are generally deficient in $F_{35}$, relative to the
 $F_{35}$ vs. $\gamma$ relation found by GB04. 
Soft events with small hard X-ray flux are missing by 
selection. This could lead to a flatter relation in reality.  

Further, it may well be possible that the power-law relation does not hold
below $F_{35}\approx10^{-5}$ photons s$^{-1}$ cm$^{-2}$ keV$^{-1}$, 
but flattens toward a constant value of $\gamma$.

The selection according to thermal flare properties aims also at a large range
of non-thermal X-ray fluxes. They range from $10^{-4}$ to $10$ photons
s$^{-1}$cm$^{-2}$keV$^{-1}$ at 35 keV ($F_{35}$, see Fig.~\ref{fluxvsgamma}), with a relatively
uniform distribution between $10^{-3}$ and 1 photons
s$^{-1}$cm$^{-2}$keV$^{-1}$. The non-thermal flux distribution depends on the
reference energy. At lower reference energy, in the range of the 'equal photon
flux', the distribution becomes narrower. 

A selection
effect comes in when the comparison includes parameters of the thermal plasma,
since among small flares events with high non-thermal flux have been
preferentially selected. The relations shown in Figs.~\ref{f35vsfgoes} and
\ref{gammavsgoesfluxsymb} may be affected
by selection. In Fig.~\ref{f35vsfgoes} small soft X-ray events with low
$F_{35}$ are missing, particularly in the interval [10$^{-5}$,10$^{-3}$]
$F_{35}$ and [10$^{-8}$,10$^{-7}$] $F_\mathrm{G}$. Thus the regression line may be
slightly flatter and the exponent in Eq.~(\ref{f35vsfgoeseq}) smaller. The trend for larger
scatter in $F_{35}$ for smaller flares may thus be even larger than visible in
Fig.~\ref{f35vsfgoes}. In Fig.~\ref{gammavsgoesfluxsymb} small soft X-ray
events with large $\gamma$ (upper left
corner) are missing. This may contribute to the absence of significant
correlation between $\gamma$ and soft X-ray flux. Note, however, that the
three intervals of non-thermal flux line up along the expected correlation
line with a large scatter below and above the
line. Nevertheless, it remains a fact that one cannot predict the spectral
index at peak non-thermal emission from the soft X-ray flux.

Different selection effects have already been mentioned in relation with the
break energy of the non-thermal spectrum at high energies, visible only in
large flares (high $F_{35}$) and low $\gamma$. We further note that
using the spectral index above the break energy, $\beta$ and its corresponding
flux normalization at 35~keV worsens the correlation. 

It may be added here also that the low-energy
turnover $E_\mathrm{turn}$ seemingly correlates with $F_{35}$ and even better with
$F_\mathrm{G}$. The latter and the fact that $E_\mathrm{turn}$ is usually close to the
energy with equal thermal and non-thermal contributions strongly point to a
spurious effect (see also Saint-Hilaire \& Benz \cite{SHilaire05}). It does not influence
significantly the relations discussed above.

The temperature changes only slightly (from about 5 MK to 30 MK) over 4 magnitudes of $F_\mathrm{GOES}$
but there is a linear dependence between $F_\mathrm{GOES}$ and
$EM_1$. Therefore, the similarity of Figs. 5 and 8 is not
surprising. In fact, the two exponents in Eqs.~(\ref{f35vsfgoeseq}) and
(\ref{emvsflux35eq}) are identical
within the statistical uncertainty. However, it is surprising that the correlation
does not improve from $F_\mathrm{G}$ vs. $F_{35}$ to $EM_1$ vs. $F_{35}$,
as $EM_1$ is determined simultaneously and at the peak of $F_{35}$ and
does not contain the temperature, whereas $F_\mathrm{G}$ is determined at the peak
of the soft X-ray emission later in the flare. $F_\mathrm{G}$ correlates well with $EM_1$ 
\begin{equation} \label{emvsfluxGOESeq}
F_\mathrm{G}=3.6\cdot 10^{-50}EM_1^{0.92},
\end{equation}
with $EM_1$ in $\mathrm{cm^{-3}}$ and $F_\mathrm{G}$ in
Wm$^{-2}$, and $b=0.92\pm 0.09$. The exponent $b$ is not significantly below
unity. Such a trend may arise from the reduced efficiency for thermal emission
at high temperatures of large flares, or may be due to larger energy losses
because of longer duration. 

The correlation between the temperature and soft X-ray emission of the thermal
 plasma (Fig. \ref{tvsgoesflux}) is steeper than previously reported by
 Feldman et al.(~\cite{Feldman96}
see Eq.~(\ref{feldmaneq})). Part of the difference can be accounted for by our subtraction
 of the background, increasing the range of peak fluxes. On the other hand,
 Feldman et al. used temperatures from the Yohkoh BCS detectors. These
 detectors saturate at high fluxes. Therefore the correlation found by
 Feldman et al. may be even
 flatter. As already mentioned in Sect.~\ref{comparetemp} RHESSI might measure
 the hotter part of a non-isothermal plasma, whereas GOES sees a cooler part.

Finally, none of the investigations on the flare position on the disk showed a
significant effect. The distributions of $\gamma$, $F_{35}$, and $E_\mathrm{br}$ are
independent of radial distance from the center of the disk. Events
  with position offset larger than 950 arcsec from the sun center have been
  looked at separately. They show no significant pattern on the $\gamma$
  vs. $F_{35}$ plot. This excludes an influence of the varying albedo of 
the X-ray emission scattered by the
chromosphere on the above results. 


\section{Summary and Conclusions}\label{summary}
Flares observed by RHESSI were selected from a wide range of GOES
flare size. The selection was nearly uniform from B1 to M6 (corresponding to
flares between A1 to M6 after background subtraction). Out of 92 flares with
a well defined spectrum at peak time, 85 events can be fitted by a thermal
and a non-thermal component. Only 6 events (7\%) had only  thermal emission,
one had only a non-thermal component. The fitted flux at 35 keV at the peak
time of the
non-thermal emission in 85 flares ranges over more than 5 orders of
magnitude (Fig.~\ref{fluxvsgamma}). This greatly exceeds the range that has been analyzed in
the temporal SHS evolution of a single subpeak (Fig. \ref{paolosflare}) and even in all
combined subpeaks in the literature (GB04). A power-law index
could be reliably determined from a minimum value of 2.3 up to a maximum of 8.3. 

The main conclusion based on this study is that the spectral index $\gamma$ at
peak non-thermal emission of flares, as well as at subpeaks of a single flare
correlate. Thus flares with small non-thermal flux are softer on the
average. This makes them even more difficult to detect in high-energy X-ray
emission. Eventually, the non-thermal emission becomes indistinguishable from
thermal emission, although it represents possibly a larger amount of energy. 

We have studied further the relations between non-thermal flux, spectral
 index, soft X-ray flux, temperature, and emission measure with the following results:
\begin{itemize}
\item The spectral index, $\gamma$, and the peak non-thermal flux at 35~keV,
 $F_{35}$, correlate linearly in a double logarithmic plot, similar to the time evolution
 of $\gamma$ and $F_{35}$ during one single flare or subpeak. 
\item Small soft X-ray emitting flares have a lower $F_{35}$ by nearly an
  order of magnitude from the relation derived earlier. They also show a larger spread in $F_{35}$. 
\item Soft X-ray flux and non-thermal flux at 35~keV are correlated,
  indicating that flares with larger non-thermal emission have larger thermal emission. 
\item Although soft X-ray flux and non-thermal flux are correlated, as well as
 spectral index and non-thermal flux, there is no significant correlation
 between spectral index and soft X-ray flux (GOES class). Probable reasons are the large
 scatter in $\gamma$ and a selection effect.
\item Temperature and emission measure of the thermal plasma both correlate
 with non-thermal flux, indicating that flares with large non-thermal flux
 have a higher temperature and emission measure. 
\item A comparison of soft X-ray flux and temperature yields a correlation also
  seen in previous work, but with slightly different relation due partially to background subtraction.
\end{itemize}

That the temperature is higher in large flares is the property of the thermal
flare plasma, thought to be heated by precipitating particles, thus a
secondary product of energy release. It may be explained by the repeated
impact of electrons accelerated in elementary flares, the larger the flare,
the more such impacts and the hotter the target. The $\gamma$ vs. $F_{35}$
relation, however, concerns the non-thermal electron population thought to be
a primary flare product. As it applies to different flares and not only to
the temporal evolution in a flare, it cannot be a secondary phenomenon. Thus
it is an intrinsic feature of the acceleration process. Its quantitative
relation must be accounted for by a realistic acceleration theory.

Lower $F_{35}$ being associated with larger $\gamma$ affect the frequency
distribution of flare energies. If the energy or peak flux of non-thermal
electrons is determined well above the mean energy of 'equal photon flux'
(10.4~keV), flares with low non-thermal flux are lost in the background. For
RHESSI observations, this affects the selection for GOES classes smaller than
about C2 (Fig.~\ref{fluxvsgamma}). The frequencies of hard X-ray flares are
reported to have power-law distributions (e.g. Hudson \cite{Hudson91}).
If soft small flares have been
missed, the published power-law indices must be considered as lower limits. 

\begin{acknowledgements}
RHESSI data analysis at ETH Z\"urich is supported by ETH grant TH-1/04-2 and the
Swiss National Science Foundation (grant 20-105366). Much of this work relied
on the RHESSI Experimental Data Center (HEDC) developed under ETH Z\"urich grant
TH-W1/99-2. We thank the many people who have contributed to the successful
operation of RHESSI and acknowledge P. Saint-Hilaire for providing the list of
events with a good image and helpful discussions.
\end{acknowledgements}

\end{document}